# On the relation between the degree of internationalization of cited and citing publications: A field level analysis, including and excluding self-citations[1]

Giovanni Abramo
*Laboratory for Studies in Research Evaluation*
*at the Institute for System Analysis and Computer Science (IASI-CNR)*
*National Research Council of Italy*
ADDRESS: Istituto di Analisi dei Sistemi e Informatica, Consiglio Nazionale delle Ricerche,
Via dei Taurini 19, 00185 Roma - ITALY
giovanni.abramo@uniroma2.it

Ciriaco Andrea D'Angelo
*University of Rome "Tor Vergata" - Italy and*
*Laboratory for Studies in Research Evaluation (IASI-CNR)*
ADDRESS: Dipartimento di Ingegneria dell'Impresa, Università degli Studi di Roma "Tor Vergata", Via del Politecnico 1, 00133 Roma - ITALY
dangelo@dii.uniroma2.it

Flavia Di Costa
*Research Value s.r.l.*
ADDRESS: Research Value, Via Michelangelo Tilli 39, 00156 Roma- ITALY
flavia.dicosta@gmail.com

**Abstract**
The growing complexity of scientific challenges demands increasingly intense research collaboration, both domestic and international. The resulting trend affects not only the modes of producing new knowledge, but also the way it is disseminated within scientific communities. This paper analyses the relationship between the "degree of internationalization" of a country's scientific production and that of the relevant citing publications. The empirical analysis is based on 2010-2012 Italian publications. Findings show: i) the probability of being cited increases with the degree of internationalization of the research team; ii) totally domestic research teams tend to cite to a greater extent totally domestic publications; iii) vice versa, publications resulting from international collaborations tend to be more cited by totally foreign publications rather than by publications including domestic authors. These results emerge both at overall and at discipline level. Findings might inform research policies geared towards internationalization.

**Keywords**
*Knowlegde flows; self-citations; bibliometrics; Italy*

# 1. Introduction

The importance of citation analysis in science is witnessed in the increasingly routine use of citational data to measure the scholarly impact of publications and scientific journals, as well as the research performance of countries, institutions, departments, and individual scientists.

The foundations of citation analysis lie in normative theory, which posits that scientists cite papers to recognize their influence (Kaplan, 1965; Merton, 1973; Martin & Irvine, 1983). The social constructivist theory objects to this view, claiming that citing to give credit is the exception, while persuasion is the main motivation (Mulkay, 1976; Bloor, 1976; Knorr-Cetina, 1991). The results from empirical testing support the normative hypothesis, and so confirm the argument that citations reflect the payment of intellectual debt (Baldi, 1998). Abramo (2018) recently revisited the relevant conceptualizations, intending to spell out some principles leading to a clear definition of the "impact" from research, and above all, of the appropriate citation-based indicator to measure it. Sociological research provides new insights, particularly concerning the role of "trusted social networks" in gathering and citing information (Thornley et al., 2015). Indeed the literature is rich in studies on the citing behavior of scientists, as seen in a review by Bornmann and Daniel (2008), and a further updated review by Tahamtan and Bornmann (2018), on the theoretical and empirical aspects of the citation process. Wouters (1999) stresses that the sciences present many types of citing culture, such that the publications within the different fields tend to share certain properties of citing: researchers in one field, for example, will tend to cite more publications than those in another (e.g. the biomedical vs. mathematics fields). "A conceptual core that is mutually shared by every one of them cannot be isolated; the various citing cultures resemble one another, as members of one family do. It is possible, of course, to abstract certain general notions and claim that these constitute the core." (Wouters, 1999).

Scholars of bibliometrics have been particularly interested in the issues of the geographical dimension of new knowledge creation (publications), i.e. the "internationalization" of research, and the spread of its impact (citations), i.e. domestic versus international spillovers, as well as the relation between these two phenomena. There is in fact an extensive literature applying bibliometric approaches to the study of international research collaboration. This includes descriptive analyses of single countries and country clusters, for China (Niu & Qiu, 2014), India (Shrivats and Bhattacharya, 2014), Italy (Abramo, D'Angelo, & Murgia, 2013), the ASEAN member states (Kumar, Rohani, & Ratnavelu, 2014), the BRICS countries (Finardi & Buratti, 2016), and the OECD countries (Choi, 2012). The US National Science Foundation's report on Science and Engineering Indicators (NSB, 2018) provides an exhaustive compendium of bibliometric data, also serving to examine the trends in international research.

Taking the normative view, in which citation linkages imply a flow of knowledge from the cited to citing authors (Mehta, Rysman, & Simcoe, 2010; Van Leeuwen & Tijssen, 2000), several scholars have relied on publication citations to investigate the international flows of scientific knowledge. Rabkin, Eisemon, Lafitte-Houssat, and McLean Rathgeber (1979) explored global visibility for four departments (botany, zoology, mathematics, and physics) of the universities of Nairobi (Kenya) and Ibadan (Nigeria). At the level of the single field, Stegmann and Grohmann (2001) measured knowledge "export" in the Dermatology & Venereal Diseases category of the 1996 CD-ROM Journal Citation Reports (JCR), and in seven unlisted dermatology journals. Hassan and Haddawy (2013) mapped knowledge flows from the United States to other countries in the "energy" field over the years 1996-2009. Abramo and D'Angelo (2018) tracked international spillovers of Italian knowledge production, in over 200 fields, by analyzing publication citations. Abramo, D'Angelo, and Carloni (2019) conceptualized the diffusion of knowledge between countries in terms of a "balance of knowledge flows" (BKF). Several studies investigated, among others, the role of geographic, cognitive and social proximity in knowledge diffusion (Jaffe, Trajtenberg, & Henderson, 1993; Hicks, Breitzman, Olivastro & Hamilton, 2001; Tijssen, 2001; Maurseth & Verspagen, 2002; Ahlgren, Persson, & Tijssen, 2013; Head, Li, & Minondo, 2018; Wuestman, Hoekman, & Frenken, 2019, Abramo, D'Angelo, & Di Costa, 2020a).

Several scholars have verified the correlation between a country's rate of international research collaboration and the impact of its publications (Bordons, Gomez, Fernandez, Zulueta, & Mendez, 1996, for Spain; Abramo, D'Angelo, & Murgia, 2017, for Italy; Kumar, Rohani, & Ratnavelu, 2014, for ASEAN nations; Kim, 2006, for Germany; Tan, Ujum, Choong, & Ratnavelu, 2015, for Malaysia). This same literature stream also verifies the existence of a "center-periphery" pattern within clusters and pairs of countries (Chinchilla-Rodríguez, Sugimoto, & Larivière, 2019; Choi, 2012; Luukkonen, Tijssen, Persson, & Sivertsen, 1993; Schubert & Sooryamoorthy, 2010).

Adams (2013) and Lancho Barrantes, Guerrero Bote, Rodríguez, and De Moya Anegón, (2012) assert that internationally coauthored papers are more highly cited because the authors are more likely to be doing excellent research. Gingras and Khelfaoui (2018) have shown the presence of a visibility (citation) advantage for the USA, given the heavy presence of American authors in bibliographic repertories such as Web of Sciences and Scopus. This fact has a knock-on effect on all those countries that collaborate more intensely with the USA.

Bornmann, Adams, and Leydesdorff (2018), analyzing the research output in the natural sciences of three economically advanced European countries (Germany, Netherlands, UK; years 2004, 2009, 2014), observe that "articles co-authored by researchers from Germany or the Netherlands are less likely to be among the globally most highly-cited articles if they also cite "domestic" research (i.e. research authored by authors from the same country)"; but this observation was not confirmed for the UK.

Fontana, Montobbio and Racca (2019) investigated how the probability of citation of papers in economics is affected by the geographic location and scientific topic of each paper. Results revealed a home bias effect in citations (for example, a publication originated in Europe is 39% more likely to get a citation from an average European publication than is a random U.S. publication).

Abramo, D'Angelo, and Di Costa (2020) investigated the relationship between the "degree of internationalization" (according to percentage of foreign authors in the byline) of 2010-2012 Italy's publications, classified in three categories (totally domestic, prevalently domestic, and prevalently international), and the "degree of internationalization" of the citing publications, also classified in three categories (totally domestic, totally international, and mixed).

In this work, we investigate nexus of the above characteristics of the byline of a paper and measure the marginal effect of the degree of internationalization of the cited publication on that of the citing publication. Furthermore, because the above relationship might be affected by self-citations (Aksnes, 2003), we investigate whether results change when self-citations are excluded.

We conduct the analysis both at overall and discipline level. This analysis allows us to answer such interesting questions as: Are a country's totally domestic publications more cited by totally domestic publications? To what extent does the probability of being cited by totally foreign publications grow as the degree of internationalization of the cited publications rises? To what extent, do self-citations affect the results? Do the results vary across fields?

The empirical analysis is based on the Italian publications of the three-year period 2010-2012 and on the citations accumulated up to 31/05/2017. The citation time window should be large enough to assure an acceptable robustness of results. The details of the dataset and methodology are illustrated in the following section. Sections 3 and 4 presents the results of the analyses, respectively including and excluding self-citations. Finally, Section 5 discusses the main findings of the work and their implications.

## 2. Data and method

Our analysis is based on the Italian-National Citation Report (I-NCR) by Clarivate Analytics, obtained by extracting all publications authored by Italian organizations from the seven main WoS core collection indexes,[2] i.e. publications with at least one affiliation showing "Italy" as country. For the 2010-2012 period, the I-NCR contains a total of 255399 records.[3] Of these, 17401 show no author-affiliation link, and are therefore excluded from the analysis. The remaining 237998 constitute the dataset and include a total of 2814990 authors, 394624 (14%) of them showing multiple affiliations.

We divide publications in the dataset into three subsets, defined as:
- "totally domestic", if all authors in the byline show Italian affiliations only;
- "prevalently domestic", if authors affiliated with Italian organizations compose less than 100%, but at least 50% of all coauthors listed in the byline;
- "prevalently international": if authors affiliated with Italian organizations compose less than 50% of all coauthors listed in the byline.

To account for multiple affiliations of a single author we adopt a fractional counting method. In case of authors with $n$ affiliations, we assign $1/n$ to each of her/his addresses. This procedure becomes critical only in the case of authors with both domestic and foreign affiliations, in then determining whether the publication is prevalently domestic or international.

The same logic could be applied to classify the citing publications, observed up to 31/05/2017, however for these the I-NCR does not indicate the author-affiliation link. Being unable to measure the prevalence of authors (Italian or foreign), we thus classify the publications into three subsets:
- "totally Italian": if the citing publication has an address list with only Italian addresses;
- "totally foreign": if the citing publications has an address list without any Italian address;

---

[2] SCI-E: Science Citation Index Expanded; SSCI: Social Sciences Citation Index; A&HCI: Arts & Humanities Citation Index; CPCI-S: Conference Proceedings Citation Index- Science; CPCI-SSH: Conference Proceedings Citation Index- Social Science & Humanities; BKCI-S: Book Citation Index– Science; BKCI-SSH: Book Citation Index– Social Sciences & Humanities.
[3] With no restrictions to any document type.

- "mixed": if the citing publications has an address list with both Italian and foreign addresses.

The following publication exemplifies the assignment to subsets:

Acocella, N., Di Bartolomeo, G., & Pauwels, W. (2010). Is there any scope for corporatism in macroeconomic policies? *Empirica*, 37(4), 403-424.

*Author-affiliation list:* [Acocella, Nicola] Univ Roma La Sapienza, Dipartimento Studi Geoecon, I-00161 Rome, Italy; [Di Bartolomeo, Giovanni] Univ Teramo, Teramo, Italy; [Pauwels, Wilfried] Univ Antwerp, B-2020 Antwerp, Belgium

The analysis of affiliations reveals that two of the three authors are affiliated with Italian organizations, so the publication is classified as prevalently domestic.
The work in question has received three citations, as follow.

Acocella, N., & Di Bartolomeo, G. (2013). The cost of social pacts. *Bulletin of Economic Research, 65*(3), 238-255.

*Address list*: Univ Roma La Sapienza, Rome, Italy; Univ Teramo, I-64100 Coste S Agostino, Italy

Schneider, F., Haigner, S. D., Jenewein, S., & Wakolbinger, F. (2014). Institutions of conflict management and economic growth revisited: A short note. *Empirica, 41*(3), 577-587.

*Address list*: Univ Linz, Dept Econ, A-4040 Linz, Austria; Univ Innsbruck, Dept Publ Econ, A-6020 Innsbruck, Austria; Gesell Angew Wirtschaftsforsch, Innsbruck, Austria

Colombo, E., Tirelli, P., & Visser, J. (2014). Reinterpreting social pacts: Theory and evidence. *Journal of Comparative Economics, 42*(2), 358-374.

*Address list*: Univ Milano Bicocca, I-20126 Milan, Italy; Univ Amsterdam, AIAS, NL-1018 TV Amsterdam, Netherlands

The first (a self-citation) presents an address list indicating only Italy (by the two authors who produced the cited publication), and therefore is classified as "totally Italian"; the second presents only Austrian addresses, and so is classified as "totally foreign"; the third presents an Italian and a Dutch address, and is classified as mixed.

To answer the research questions, we apply two OLS models, where the response variables are, respectively, number of citations by totally foreign publications and by totally Italian publications. The formula of the model is:

$$Y = b_0 + b_1 X_1 + b_2 X_2 + b_3 X_3 + b_4 X_4 + b_5 X_5 + b_6 X_6 + b_7 X_7$$

[1]

Where:
$Y = \ln(1 + \text{citations by totally foreign/totally Italian publications})$
$X_1$ = Number of authors of the cited publication
$X_2$ = Total citations received
$X_3$ = Prevalently domestic cited publication
$X_4$ = Prevalently international cited publication
$X_5$ = Review
$X_6$ = Proceeding
$X_7$ = Other document types

We adopt a log transformation because of the highly skewed citation distributions, which are problematic for OLS regressions. We add one to allow for inclusion of uncited publications.

The independent variable $X_1$ allows to control for high number of co-authors, which often occurs in such fields as particle physics, oceanography, climate change, clinical medicine, assuming that papers by large international research teams present wider interest for the global scientific community. The independent variable $X_2$ allows to capture the effect of the degree of internationalization of the cited publication on that of the citing ones, with impact being equal. Binary variables $X_3$ and $X_4$ are those under study. Their combination defines the byline type (totally domestic, prevalently domestic, prevalently international) of the cited publication, assuming as baseline "totally domestic" bylines.

The three dummy variables refer to the cited publication document types (assuming as baseline the document type "article"), because the intensity of citations received depend, all others equal, on document type.

The following sections present the results of the analyses. After presenting the data at the overall level, we detail the analysis at the level of the macro-scientific domain (disciplinary area). To do this, each publication is first assigned to the subject category (SC) of its hosting journal, conference or book,[4] as per the WoS classification scheme, and so to the relative macro-disciplinary area.[5] A final subsection will show the results after excluding self-citations.

## 3. Results

### 3.1 Overall level analysis

Out of the dataset of 233998 publications, 138313 (58.1%) are totally domestic, 46433 (19.5%) are prevalently domestic, and 53252 (22.4%) are prevalently international (Table 1). The share of publications cited is directly related to the degree of internationalization: the prevalently international ones are cited in 82.3% of cases, prevalently and totally domestic ones in respectively 80.4% and 68.2% of cases. Altogether the publications of the dataset received 3282334 citations, of which almost three quarters were from totally foreign publications, while the remainder were almost equally from totally Italian (13.1%) or mixed (12.5%) ones.

However, this distribution of citing publications differs considerably in relation to the degree of internationalization of the cited publication.

The totally domestic Italian publications receive only 67.0% of their citations from totally foreign publications, compared to 70.3% for the prevalently domestic and 82.9% for the prevalently international ones. The correlation between the degree of internationalization of the cited and citing publications is also evident also considering the other cells of the matrix: totally domestic products receive 22.7% of their total citations from totally Italian publications, against 12.7% for the prevalently domestic and only 4.9% for the prevalently international ones.

---

[4] Publications hosted in multi-category sources are assigned to each of the SCs.
[5] Our assignment of SCs to disciplinary areas (Mathematics; Physics; Chemistry; Earth and Space Sciences; Biology; Biomedical Research; Psychology; Clinical Medicine; Engineering; Economics; Law, political and social sciences) follows a pattern previously published in the ISI Journal Citation Reports website, although this information is no longer available through the Clarivate web portal.

The Pearson's $\chi^2$ value observed (19.6E+4 with p-value 0.000) demonstrates the significance of the association between the degree of internationalization of Italian publications and that of the relative citing publications. Cramer's V test, with control for sample size, provides a further measure of the strength of the association and its practical significance: in fact the value (0.173) indicates a rather weak but still significant association between the two variables.

*Table 1: Distribution of 2010-2012 Italian publications in function of their degree of internationalization and the relative citing publications*

| Category of Italian publications | No. of Italian publications | Of which cited | Total citations | Of which totally Italian | Of which mixed | Of which totally foreign |
|---|---|---|---|---|---|---|
| Totally domestic | 138313 (58.1%) | 68.2% | 1227947 (37.4%) | 22.7% | 10.3% | 67.0% |
| Prevalently domestic | 46433 (19.5%) | 80.4% | 670833 (20.4%) | 12.7% | 17.0% | 70.3% |
| Prevalently international | 53252 (22.4%) | 82.3% | 1383554 (42.2%) | 4.9% | 12.2% | 82.9% |
| Total | 237998 | 73.7% | 3282334 | 13.1% | 12.5% | 74.4% |

Next we show the outcomes of the OLS regressions. Table 2 shows the coefficients of the model predicting the number of citations that Italian publications receive by totally Italian publications. Because of the log transformation, the marginal effect is represented by the exponential of the coefficient. With respect to a totally domestic publication, a prevalently domestic is cited 13.4% ($e^{-0.144}$) less by totally domestic publications, and a prevalently international 34.3% ($e^{-0.420}$) less, with total citations, number of authors, and document type being equal.

Table 3 shows the coefficients of the model predicting the number of citations that Italian publications receive by totally foreign publications. In this case, the model shows that, with respect to a totally domestic publication, a prevalently domestic is cited 25.7% ($e^{0.229}$) more, and a prevalently international 74.4% ($e^{0.556}$), more, all others being equal.

In both models, the control variables play a role, remarkable in the case of the document type, and less so in the case of the number of authors and total citations.

*Table 2: OLS regression for 2010-2012 Italian publications. Y = Number of citations that Italian publications receive from totally Italian publications (Log transformed).*

|  | Coeff. | Std Err. | t | P>\|t\| | [95% Conf. Interval] | |
|---|---|---|---|---|---|---|
| Const | 0.830 | 0.008 | 108.2 | 0.000 | 0.815 | 0.845 |
| Number of authors | 0.000 | 0.000 | -5.5 | 0.000 | 0.000 | 0.000 |
| Total_citations | 0.006 | 0.001 | 11.3 | 0.000 | 0.005 | 0.008 |
| Cited_pub_prevalently_domestic | -0.144 | 0.004 | -33.1 | 0.000 | -0.153 | -0.136 |
| Cited_pub_prevalently_international | -0.420 | 0.009 | -48.5 | 0.000 | -0.436 | -0.403 |
| Review | 0.022 | 0.009 | 2.4 | 0.019 | 0.004 | 0.041 |
| Proceeding | -0.554 | 0.008 | -65.6 | 0.000 | -0.570 | -0.537 |
| Other doc. types | -0.660 | 0.009 | -74.6 | 0.000 | -0.678 | -0.643 |

Baseline: "Cited publication: totally_domestic"; "articles" (as document type).
Number of obs = 237997
F( 7,237989) =14637.5; Prob > F = 0.0000
R-squared = 0.2847; Root MSE = .67526

*Table 3: OLS regression for 2010-2012 Italian publications. Y = Number of citations that Italian publications receive from totally foreign publications (Log transformed).*

|  | Coeff. | Std Err. | t | P>|t| | [95% Conf. Interval] | |
|---|---|---|---|---|---|---|
| Const | 1.428 | 0.015 | 96.2 | 0.000 | 1.398 | 1.457 |
| Number of authors | 0.000 | 0.000 | 1.7 | 0.086 | 0.000 | 0.000 |
| Total_citations | 0.012 | 0.001 | 10.7 | 0.000 | 0.010 | 0.015 |
| Cited_pub_prevalently_domestic | 0.229 | 0.007 | 34.0 | 0.000 | 0.216 | 0.242 |
| Cited_pub_prevalently_international | 0.556 | 0.017 | 33.3 | 0.000 | 0.523 | 0.589 |
| Review | 0.281 | 0.016 | 17.4 | 0.000 | 0.249 | 0.312 |
| Proceeding | -1.221 | 0.016 | -76.3 | 0.000 | -1.252 | -1.189 |
| Other doc. types | -1.389 | 0.017 | -80.1 | 0.000 | -1.423 | -1.355 |

*Baseline: "Cited publication: totally_domestic"; "articles" (as document type).*
Number of obs = 237997
F( 7,237989) =42801.2; Prob > F = 0.0000
R-squared = 0.5069; Root MSE = .91606

### 3.2 Variability across disciplinary areas

Both intensity of publication (D'Angelo & Abramo, 2015) and citation behavior are known to vary across disciplines. We could then expect that the degree of internationalization of scientific production and that of the relevant citing publications would also vary across disciplines. For this reason we repeat the analysis just presented, at the level of single disciplinary area.

As shown in Table 4, in Art and Humanities, three publications out of every four are authored by scholars who are all affiliated only with Italian institutions. In Biomedical research, Clinical medicine and Engineering, the incidence of totally domestic publications is higher than 60%; in all the other areas it is higher than 50%, with the sole exception of Physics (43.4%). On the other hand, in Art and Humanities, only 9.4% of publications are prevalently international. Values of less than 20% are also recorded in Mathematics (17.1%) and Engineering (17.7%). Physics again represents the exception, showing a peak of 31.9% of authored publications by prevalently international teams, followed by Earth and space sciences (25.4%) and Chemistry (21.9%).

With regard to the incidence of cited publications out of the total, we observe an increasing trend with the degree of internationalization, meaning in progressing from totally to prevalently domestic to prevalently international. This is observed in all disciplinary areas except Art and humanities, Biomedical research, Economics, and Law, political and social sciences - areas whose contents are evidently more dependent on the national context. Regardless of the degree of internationalization of the Italian publications, the citations come mainly from "totally foreign" citing articles. The highest value is found in Clinical medicine (78.4% at the overall level) but percentages higher than 75% are also observed in other biomedical areas. The "totally Italian" citing papers more often refer to "domestic" papers than "international" ones: in Earth and space sciences the incidence of the totally Italian citing publications drops from 30.8% for totally domestic publications to 4.8% for prevalently international ones. Conversely, the incidence of totally foreign citing publications increases with the level of internationalization of the Italian publications in reference: in Earth and space sciences there is again a difference in incidence of almost 30 percentage points (53.4% for totally domestic vs. 81.8% for prevalently domestic).

The tests of association between the two nominal variables (category of cited and citing publications) are reported in Table 5. The calculations of Pearson $\chi^2$ and corresponding p-value indicate a significant association; the values of Cramer's V, never higher than 0.25, reveal that the association is moderate in Art and Humanities, Earth and space sciences and Physics, but otherwise almost always weak, with the minimum value observed in Biomedical research.

*Table 4: Distribution of 2010-2012 Italian publications in function of their degree of internationalization and the relative citing publications, by disciplinary area*

| Disciplinary area | Category of Italian publication | No. of Italian publications | Of which cited (%) | Total citations | Of which totally Italian (%) | Of which mixed (%) | Of which totally foreign (%) |
|---|---|---|---|---|---|---|---|
| Art and humanities | Totally domestic | 2491 (75.1%) | 20.4 | 3331 (55.4%) | 35.7 | 13.2 | 51.1 |
| | Prevalently domestic | 516 (15.5%) | 46.7 | 1447 (24.0%) | 16.9 | 17.5 | 65.6 |
| | Prevalently international | 312 (9.4%) | 46.5 | 1240 (20.6%) | 7.6 | 12.0 | 80.4 |
| | Total | 3319 | 26.9 | 6018 | 25.4 | 14.0 | 60.6 |
| Biology | Totally domestic | 22044 (57.5%) | 77.7 | 244158 (41.7%) | 22.8 | 10.0 | 67.2 |
| | Prevalently domestic | 8087 (21.1%) | 83.7 | 131362 (22.5%) | 13.0 | 15.0 | 72.0 |
| | Prevalently international | 8192 (21.4%) | 87.7 | 209432 (35.8%) | 4.6 | 9.4 | 85.9 |
| | Total | 38323 | 81.1 | 584952 | 14.1 | 10.9 | 75.0 |
| Biomedical research | Totally domestic | 27283 (62.2%) | 65.3 | 263432 (39.9%) | 19.8 | 8.7 | 71.4 |
| | Prevalently domestic | 7053 (16.1%) | 77.5 | 124481 (18.9%) | 12.8 | 13.6 | 73.6 |
| | Prevalently international | 9541 (21.7%) | 76.2 | 271704 (41.2%) | 6.1 | 9.0 | 85.0 |
| | Total | 43877 | 69.7 | 659617 | 12.8 | 9.8 | 77.4 |
| Chemistry | Totally domestic | 9777 (54.9%) | 93.2 | 160606 (46.4%) | 22.1 | 9.5 | 68.4 |
| | Prevalently domestic | 4121 (23.2%) | 94.1 | 84989 (24.5%) | 12.6 | 15.7 | 71.7 |
| | Prevalently international | 3901 (21.9%) | 94.5 | 100649 (29.1%) | 4.3 | 10.2 | 85.5 |
| | Total | 17799 | 93.7 | 346244 | 14.6 | 11.2 | 74.2 |
| Clinical medicine | Totally domestic | 48525 (63.5%) | 63.4 | 399536 (37.5%) | 20.2 | 8.4 | 71.4 |
| | Prevalently domestic | 11457 (15.0%) | 75.4 | 172282 (16.1%) | 12.4 | 13.8 | 73.8 |
| | Prevalently international | 16402 (21.5%) | 77.5 | 495003 (46.4%) | 5.3 | 9.2 | 85.6 |
| | Total | 76384 | 68.2 | 1066821 | 12.0 | 9.6 | 78.4 |
| Earth and space sciences | Totally domestic | 7262 (51.3%) | 86.5 | 86476 (37.7%) | 30.8 | 15.8 | 53.4 |
| | Prevalently domestic | 3311 (23.4%) | 91.6 | 52623 (22.9%) | 15.9 | 21.6 | 62.5 |
| | Prevalently international | 3595 (25.4%) | 92.5 | 90294 (39.4%) | 4.8 | 13.4 | 81.8 |
| | Total | 14168 | 89.2 | 229393 | 17.1 | 16.2 | 66.7 |
| Economics | Totally domestic | 3128 (53.7%) | 58.7 | 15663 (39.3%) | 21.5 | 10.2 | 68.3 |
| | Prevalently domestic | 1504 (25.8%) | 73.7 | 12892 (32.4%) | 8.6 | 12.2 | 79.1 |
| | Prevalently international | 1190 (20.4%) | 71.3 | 11250 (28.3%) | 5.4 | 10.8 | 83.8 |
| | Total | 5822 | 65.1 | 39805 | 12.8 | 11.0 | 76.2 |
| Engineering | Totally domestic | 29276 (62.2%) | 68.7 | 225875 (49.8%) | 26.1 | 10.3 | 63.7 |
| | Prevalently domestic | 9499 (20.2%) | 76.7 | 104463 (23.0%) | 13.6 | 16.2 | 70.2 |
| | Prevalently international | 8320 (17.7%) | 78.0 | 123441 (27.2%) | 5.5 | 11.1 | 83.4 |
| | Total | 47095 | 71.9 | 453779 | 17.6 | 11.9 | 70.5 |
| Law, political and social sciences | Totally domestic | 3299 (58.5%) | 48.3 | 12675 (40.4%) | 25.8 | 10.2 | 64.0 |
| | Prevalently domestic | 1213 (21.5%) | 66.8 | 7815 (24.9%) | 10.6 | 14.4 | 75.0 |
| | Prevalently international | 1130 (20.0%) | 66.5 | 10860 (34.6%) | 4.3 | 11.5 | 84.2 |
| | Total | 5642 | 55.9 | 31350 | 14.5 | 11.7 | 73.8 |
| Mathematics | Totally domestic | 4664 (50.4%) | 74.3 | 27418 (42.2%) | 31.2 | 15.2 | 53.6 |
| | Prevalently domestic | 3011 (32.5%) | 82.6 | 22968 (35.3%) | 13.9 | 23.1 | 63.0 |
| | Prevalently international | 1579 (17.1%) | 83.1 | 14641 (22.5%) | 6.9 | 15.9 | 77.2 |
| | Total | 9254 | 78.5 | 65027 | 19.6 | 18.1 | 62.2 |
| Physics | Totally domestic | 16069 (43.4%) | 75.2 | 156229 (25.6%) | 24.3 | 14.6 | 61.1 |
| | Prevalently domestic | 9124 (24.6%) | 86.1 | 135337 (22.2%) | 11.7 | 24.4 | 63.9 |
|  | Prevalently international | 11827 (31.9%) | 89.6 | 317703 (52.1%) | 3.6 | 20.8 | 75.6 |
|  | Total | 37020 | 82.5 | 609269 | 10.7 | 20.0 | 69.3 |
| Psychology | Totally domestic | 2088 (57.3%) | 62.0 | 15468 (37.9%) | 24.2 | 11.9 | 63.9 |
|  | Prevalently domestic | 823 (22.6%) | 80.0 | 10517 (25.8%) | 12.9 | 16.3 | 70.8 |
|  | Prevalently international | 733 (20.1%) | 86.6 | 14837 (36.3%) | 5.0 | 10.0 | 85.0 |
|  | Total | 3644 | 71.0 | 40822 | 14.3 | 12.3 | 73.4 |

*Table 5: Tests\* of association between the degree of internationalization of Italian publications and that of the relative citing publications, by disciplinary area*

| Disciplinary area | Obs | Pearson $\chi^2$ | p-value | Cramer's V |
|---|---:|---:|---:|---:|
| Art and humanities | 3319 | 0.5E+03 | 0.000 | 0.202 |
| Biology | 38323 | 3.5E+04 | 0.000 | 0.172 |
| Biomedical research | 43877 | 2.6E+04 | 0.000 | 0.139 |
| Chemistry | 17799 | 1.8E+04 | 0.000 | 0.162 |
| Clinical medicine | 76384 | 5.1E+04 | 0.000 | 0.155 |
| Earth and space sciences | 14168 | 2.4E+04 | 0.000 | 0.231 |
| Economics | 5822 | 1.8E+03 | 0.000 | 0.152 |
| Engineering | 47095 | 2.7E+04 | 0.000 | 0.172 |
| Law, political and social sciences | 5642 | 2.4E+03 | 0.000 | 0.194 |
| Mathematics | 9254 | 4.8E+03 | 0.000 | 0.192 |
| Physics | 37020 | 5.0E+04 | 0.000 | 0.202 |
| Psychology | 3644 | 2.7E+03 | 0.000 | 0.180 |

\* *(three categories of both cited publications and citing publications) => 3x3 contingency matrices = 4 degree of freedom*

As for the marginal effect of the degree of internationalization of the cited publication on that of the citing, Table 6 and Table 7 show the results of the OLS regressions. Findings align with those at overall levels, although variabilities across disciplines occur. In Table 6, columns 3 and 4 show that, all others being equal, a prevalently domestic publication receives from totally domestic publications 3% to 26% fewer citations than a totally domestic, and a prevalently international 23% to 57% fewer citations. The marginal effect, statistically significant in all disciplines, is outstanding in Chemistry and in Earth and space sciences, and remarkable in Mathematics and in Physics.

As for citations that Italian publications receive from totally foreign publications, Table 7 show that the effects of the degree of internationalization of cited publications is statistically significant in all disciplines. All others equal, as compared to totally domestic publications, prevalently domestic ones receive from totally foreign publications from a minimum of 12% citations more in Art and humanities to a maximum of 32% in Clinical medicine. Prevalently international publications receive 84% more citations than totally domestic in Earth and space sciences, +82% in Physics, +79% in Clinical medicine.

*Table 6: OLS regression for 2010-2012 Italian publications. Y = Number of citations that Italian publications receive from totally Italian publications (Log transformed), by disciplinary area.*

| Area | Number of obs | Cited_pub_prevalently_domestic | Cited_pub_prevalently_international | F | Prob > F | Root MSE | R-squared |
|---|---|---|---|---|---|---|---|
| Art and humanities | 3319 | -0.166*** | -0.372*** | 142.9 | 0.000 | 0.328 | 0.547 |
| Biology | 38323 | -0.190*** | -0.585*** | 3070.6 | 0.000 | 0.668 | 0.336 |
| Biomedical research | 43877 | -0.032*** | -0.314*** | 4455.6 | 0.000 | 0.640 | 0.386 |
| Chemistry | 17799 | -0.325*** | -0.831*** | 905.7 | 0.000 | 0.730 | 0.274 |
| Clinical medicine | 76384 | -0.028*** | -0.256*** | 5900.6 | 0.000 | 0.641 | 0.330 |
| Earth and space sciences | 14168 | -0.344*** | -0.846*** | 793.6 | 0.000 | 0.726 | 0.310 |
| Economics | 5822 | -0.254*** | -0.373*** | 162.0 | 0.000 | 0.498 | 0.333 |
| Engineering | 47095 | -0.239*** | -0.511*** | 2202.6 | 0.000 | 0.656 | 0.303 |
| Law, political and social sciences | 5642 | -0.230*** | -0.420*** | 227.5 | 0.000 | 0.478 | 0.368 |
| Mathematics | 9254 | -0.334*** | -0.541*** | 246.7 | 0.000 | 0.609 | 0.265 |
| Physics | 37020 | -0.284*** | -0.598*** | 1487.7 | 0.000 | 0.688 | 0.238 |
| Psychology | 3644 | -0.200*** | -0.519*** | 254.2 | 0.000 | 0.635 | 0.318 |

*Baseline: "Cited publication totally_domestic". Controlling for total citations, number of authors and document type.*

*Table 7: OLS regression for 2010-2012 Italian publications. Y = Number of citations that Italian publications receive from totally foreign publications (Log transformed), by disciplinary area.*

| Area | Number of obs | Cited_pub_prevalently_domestic | Cited_pub_prevalently_international | F | Prob > F | Root MSE | R-squared |
|---|---|---|---|---|---|---|---|
| Art and humanities | 3319 | 0.116*** | 0.123*** | 379.0 | 0.000 | 0.341 | 0.759 |
| Biology | 38323 | 0.212*** | 0.499*** | 9167.6 | 0.000 | 0.827 | 0.578 |
| Biomedical research | 43877 | 0.252*** | 0.477*** | 13513.9 | 0.000 | 0.838 | 0.624 |
| Chemistry | 17799 | 0.163*** | 0.357*** | 1189.4 | 0.000 | 0.862 | 0.440 |
| Clinical medicine | 76384 | 0.278*** | 0.583*** | 19419.8 | 0.000 | 0.876 | 0.580 |
| Earth and space sciences | 14168 | 0.261*** | 0.612*** | 1827.7 | 0.000 | 0.828 | 0.536 |
| Economics | 5822 | 0.142*** | 0.214*** | 665.5 | 0.000 | 0.662 | 0.661 |
| Engineering | 47095 | 0.155*** | 0.330*** | 5943.4 | 0.000 | 0.834 | 0.513 |
| Law, political and social sciences | 5642 | 0.166*** | 0.267*** | 815.5 | 0.000 | 0.629 | 0.659 |
| Mathematics | 9254 | 0.199*** | 0.365*** | 537.1 | 0.000 | 0.745 | 0.471 |
| Physics | 37020 | 0.200*** | 0.601*** | 5678.1 | 0.000 | 0.970 | 0.429 |
| Psychology | 3644 | 0.213*** | 0.411*** | 1305.1 | 0.000 | 0.758 | 0.634 |

*Baseline: "Cited publication totally_domestic". Controlling for total citations, number of authors and document type*

### 3.3 The role of self-citations

In the above analysis, we have included the count of self-citations.[6] Because a publication self-citing a totally domestic publication is by definition domestic as well, and the probability that a publication self-citing a prevalently international publication be international is higher, including or excluding self-citations is likely to affect the results of the analysis. We then replicate the same procedure as in Section 3, for the dataset without self-citations.

The descriptive statistics of the new dataset are shown in Table 8 at overall level. Total citations are now just short of 2.5 million, as 24% of citations shown in Table 1 were self-citations. Comparing with Table 1, the distribution of cited publications across categories does not change significantly: 37.1% are totally domestic publications (vs 37.4, including self-citations); 20.5% are prevalently domestic (vs 20.4); 42.3% are prevalently international (vs 42.2). Notable changes occur instead in the distribution of citing publications: 88.6% of total citations now come from totally foreign publications (vs 74.4, including self-citations), 7.8% from totally Italian (vs 13.1%) and 3.6% from mixed ones (vs 12.5%).

Totally domestic publications register the most conspicuous variations: the proportion of citations from totally foreign publications shows a 21% increase (88.3% vs 67.0%), while that from "totally Italian" and from mixed diminishes (11.7% vs 22.7% and 0.1% vs 10.3%, respectively).

For the "prevalently international" publications, the proportion of citations from "totally foreign" publications raises (90.2% vs 82.9%), the one from "totally Italian" remains stable, while the one from mixed diminishes (3.6% vs 12.2%).

Table 9 shows the relation between the categories of cited publications and those of citing (excluding self-citations), at disciplinary level. Comparing with Table 4 (with self-citations), one observes the same variations as those at the overall level, whereby those involving Mathematics and Chemistry are relevant.

Table 10 shows the results of the tests of association between the two nominal variables (category of cited and citing publications). The calculations of Pearson $\chi^2$ and corresponding p-value indicate a significant association; the values of Cramer's V vary between 0.17 (Art and Humanities) and 0.04 (Biomedical research). Differences between these values and the corresponding ones in Table 5 are invariably negative, meaning a substantial reduction of the degree of association between the two nominal variables (category of cited and category of citing publication). Highest variations occur in Chemistry, Clinical medicine and Biomedical research, lowest in Engineering, Economics and Art and humanities.

---

[6] A "self-citation" is said to occur here, when the citing and cited publications show at least one author in common.

*Table 8: Distribution of 2010-2012 Italian publications in function of their degree of internationalization and the relative citing publications (without self-citations)*

| Category of Italian publications | No. of Italian publications | Of which cited | Total citations | Of which totally Italian | Of which mixed | Of which totally foreign |
|---|---|---|---|---|---|---|
| Totally domestic | 138313 (58.1%) | 65.3% | 928402 (37.1%) | 11.7% | 0.1% | 88.3% |
| Prevalently domestic | 46433 (19.5%) | 77.1% | 513638 (20.5%) | 7.4% | 6.7% | 85.9% |
| Prevalently international | 53252 (22.4%) | 79.6% | 1057566 (42.3%) | 4.5% | 5.3% | 90.2% |
| Total | 237998 | 70.8% | 2499606 | 7.8% | 3.6% | 88.6% |

*Table 9: Distribution of 2010-2012 Italian publications in function of their degree of internationalization and the relative citing publications (without self-citations), by disciplinary area*

| Disciplinary area | Category of Italian publication | No. of Italian publications | Of which cited (%) | Total citations | Of which totally Italian (%) | Of which mixed (%) | Of which totally foreign (%) |
|---|---|---|---|---|---|---|---|
| Art and Humanities | Totally_domestic | 2491 (75.1%) | 9.3 | 2577 (55.7%) | 26.0 | 8.8 | 65.2 |
|  | Prevalently_domestic | 516 (15.5%) | 24.0 | 1097 (23.7%) | 13.2 | 7.0 | 79.8 |
|  | Prevalently_international | 312 (9.4%) | 24.7 | 951 (20.6%) | 5.6 | 4.9 | 89.5 |
|  | Total | 3319 | 13.0 | 4625 | 18.8 | 7.6 | 73.6 |
| Biology | Totally_domestic | 22044 (57.5%) | 56.0 | 192453 (42.3%) | 10.0 | 5.0 | 85.0 |
|  | Prevalently_domestic | 8087 (21.1%) | 68.5 | 100443 (22.1%) | 6.7 | 5.0 | 88.3 |
|  | Prevalently_international | 8192 (21.4%) | 74.2 | 162471 (35.7%) | 3.9 | 4.1 | 92.0 |
|  | Total | 38323 | 62.5 | 455367 | 7.1 | 4.7 | 88.2 |
| Biomedical Research | Totally_domestic | 27283 (62.2%) | 42.3 | 216399 (40.7%) | 8.5 | 4.8 | 86.7 |
|  | Prevalently_domestic | 7053 (16.1%) | 61.6 | 97964 (18.4%) | 6.5 | 4.6 | 88.9 |
|  | Prevalently_international | 9541 (21.7%) | 63.0 | 216822 (40.8%) | 5.4 | 3.9 | 90.7 |
|  | Total | 43877 | 49.9 | 531185 | 6.9 | 4.4 | 88.7 |
| Chemistry | Totally_domestic | 9777 (54.9%) | 74.7 | 122964 (46.7%) | 7.4 | 3.6 | 89.0 |
|  | Prevalently_domestic | 4121 (23.2%) | 81.8 | 62648 (23.8%) | 5.3 | 4.2 | 90.5 |
|  | Prevalently_international | 3901 (21.9%) | 82.7 | 77474 (29.4%) | 2.9 | 3.3 | 93.8 |
|  | Total | 17799 | 78.1 | 263086 | 5.6 | 3.7 | 90.8 |
| Clinical Medicine | Totally_domestic | 48525 (63.5%) | 37.4 | 331354 (38.6%) | 9.2 | 5.0 | 85.9 |
|  | Prevalently_domestic | 11457 (15.0%) | 56.6 | 137201 (16.0%) | 7.1 | 5.2 | 87.7 |
|  | Prevalently_international | 16402 (21.5%) | 64.0 | 390770 (45.5%) | 4.9 | 4.1 | 91 |
|  | Total | 76384 | 46.0 | 859325 | 6.9 | 4.6 | 88.5 |
| Earth and Space Sciences | Totally_domestic | 7262 (51.3%) | 64.5 | 65043 (38.3%) | 19.2 | 10.1 | 70.7 |
|  | Prevalently_domestic | 3311 (23.4%) | 75.5 | 38639 (22.8%) | 11.6 | 10.4 | 78.0 |
|  | Prevalently_international | 3595 (25.4%) | 79.0 | 66134 (38.9%) | 4.4 | 6.7 | 88.9 |
|  | Total | 14168 | 70.7 | 169816 | 11.7 | 8.8 | 79.5 |
| Economics | Totally_domestic | 3128 (53.7%) | 27.3 | 13446 (39.0%) | 13.7 | 7.1 | 79.2 |
|  | Prevalently_domestic | 1504 (25.8%) | 39.8 | 11392 (33.0%) | 6.8 | 7.4 | 85.8 |
|  | Prevalently_international | 1190 (20.4%) | 40.3 | 9655 (28.0%) | 4.9 | 6.2 | 88.9 |
|  | Total | 5822 | 33.2 | 34493 | 8.9 | 7.0 | 84.1 |
| Engineering | Totally_domestic | 29276 (62.2%) | 36.0 | 183523 (50.4%) | 16.0 | 6.0 | 78.0 |
|  | Prevalently_domestic | 9499 (20.2%) | 49.3 | 81506 (22.4%) | 8.6 | 7.1 | 84.3 |
|  | Prevalently_international | 8320 (17.7%) | 50.7 | 99038 (27.2%) | 4.7 | 5.1 | 90.2 |
|  | Total | 47095 | 41.3 | 364067 | 11.3 | 6.0 | 82.7 |
| Law, political and social sciences | Totally_domestic | 3299 (58.5%) | 22.9 | 10337 (40.5%) | 15.4 | 6.6 | 78.0 |
|  | Prevalently_domestic | 1213 (21.5%) | 33.8 | 6466 (25.4%) | 8.4 | 6.3 | 85.4 |
|  | Prevalently_international | 1130 (20.0%) | 38.8 | 8693 (34.1%) | 3.6 | 4.5 | 91.9 |
|  | Total | 5642 | 28.4 | 25496 | 9.6 | 5.8 | 84.6 |

| Disciplinary area | Category of Italian publication | No. of Italian publications | Of which cited (%) | Total citations | Of which totally Italian (%) | Of which mixed (%) | Of which totally foreign (%) |
|---|---|---|---|---|---|---|---|
| Mathematics | Totally_domestic | 4664 (50.4%) | 50.4 | 19662 (42.2%) | 16.9 | 9.2 | 73.9 |
| | Prevalently_domestic | 3011 (32.5%) | 61.8 | 16065 (34.5%) | 10.2 | 10.1 | 79.7 |
| | Prevalently_international | 1579 (17.1%) | 59.8 | 10872 (23.3%) | 5.9 | 7.3 | 86.8 |
| | Total | 9254 | 55.7 | 46599 | 12.0 | 9.1 | 78.9 |
| Physics | Totally_domestic | 16069 (43.4%) | 50.9 | 117876 (27.2%) | 12.4 | 7.6 | 80.1 |
| | Prevalently_domestic | 9124 (24.6%) | 69.8 | 96554 (22.3%) | 7.3 | 10.9 | 81.8 |
| | Prevalently_international | 11827 (31.9%) | 77.2 | 218684 (50.5%) | 3.6 | 9.0 | 87.4 |
| | Total | 37020 | 64.0 | 433114 | 6.8 | 9.0 | 84.2 |
| Psychology | Totally_domestic | 2088 (57.3%) | 42.7 | 11798 (37.6%) | 10.5 | 6.2 | 83.3 |
| | Prevalently_domestic | 823 (22.6%) | 64.3 | 7877 (25.1%) | 7.7 | 6.4 | 85.9 |
| | Prevalently_international | 733 (20.1%) | 70.4 | 11710 (37.3%) | 4.1 | 4.2 | 91.7 |
| | Total | 3644 | 53.2 | 31385 | 7.4 | 5.5 | 87.1 |

*Table 10: Tests\* of association between the degree of internationalization of Italian publications and that of the relative citing publications (without self-citations), by disciplinary area*

| Disciplinary area | Publications | Pearson $\chi^2$ | p-value | Cramer's V |
|---|---|---|---|---|
| Art and humanities | 3319 | 2.55E+02 | 0.000 | 0.166 |
| Biology | 38323 | 5.37E+03 | 0.000 | 0.077 |
| Biomedical research | 43877 | 1.89E+03 | 0.000 | 0.042 |
| Chemistry | 17799 | 1.89E+03 | 0.000 | 0.060 |
| Clinical medicine | 76384 | 5.70E+03 | 0.000 | 0.058 |
| Earth and space sciences | 14168 | 8.07E+03 | 0.000 | 0.154 |
| Economics | 5822 | 6.53E+02 | 0.000 | 0.097 |
| Engineering | 47095 | 9.53E+03 | 0.000 | 0.114 |
| Law political and social sciences | 5642 | 8.54E+02 | 0.000 | 0.129 |
| Mathematics | 9254 | 9.84E+02 | 0.000 | 0.103 |
| Physics | 37020 | 1.00E+04 | 0.000 | 0.108 |
| Psychology | 3644 | 4.34E+02 | 0.000 | 0.083 |

\* *(three categories of both the cited publications and the citing publications) => 3x3 contingency matrices = 4 degree of freedom*

Table 11 and Table 12 show the results of the OLS regressions carried out on the dataset without self-citations. As for the number of citations that Italian publications receive by totally Italian publications, at the overall level the marginal effects are much lower than when including self-citations (Table 2). The last row of Table 11 show that a prevalently domestic publication receives from totally domestic publications 4% fewer citations than a totally domestic, and a prevalently international 10% fewer.

At discipline level, variability is slight. In general the marginal effects are statistically significant. In Biomedical research the $X_3$ coefficient, though not significant, is surprisingly positive. In Clinical medicine, the coefficient is positive and significant. Earth and space sciences shows the highest marginal effects: a prevalently domestic publication receives from totally domestic publications 14% fewer citations than a totally domestic, and a prevalently international, 34% fewer.

Considering the number of citations received from totally foreign publications, Table 12 reveals marginal effects significant and aligned in all disciplines, although less strong than when including self-citations (Table 3). At overall level, the last row of Table 12 shows that with respect to a totally domestic publication, a prevalently domestic is cited 21% more by totally foreign publications, and a prevalently international 52%, more, all others being equal.

*Table 11: OLS regression for 2010-2012 Italian publications. Y = Number of citations that Italian publications receive from totally Italian publications (Log transformed), by disciplinary area, excluding self-citations.*

| Area | Number of obs. | Cited_pub_prevalently _domestic | Cited_pub_prevalently _international | F | Prob > F | Root MSE | R-squared |
|---|---|---|---|---|---|---|---|
| Art and humanities | 3319 | -0.107*** | -0.244*** | 86.7 | 0 | 0.278 | 0.442 |
| Biology | 38323 | -0.058*** | -0.171*** | 1569.4 | 0 | 0.512 | 0.247 |
| Biomedical research | 43877 | 0.004 | -0.019* | 2516.4 | 0 | 0.479 | 0.344 |
| Chemistry | 17799 | -0.069*** | -0.203*** | 257.4 | 0 | 0.531 | 0.156 |
| Clinical medicine | 76384 | 0.017*** | -0.007 | 3701.1 | 0 | 0.480 | 0.305 |
| Earth and space sciences | 14171 | -0.153*** | -0.416*** | 269.0 | 0 | 0.637 | 0.192 |
| Economics | 5822 | -0.100*** | -0.174*** | 94.0 | 0 | 0.416 | 0.285 |
| Engineering | 47095 | -0.130*** | -0.247*** | 601.7 | 0 | 0.555 | 0.160 |
| Law political and social sciences | 5642 | -0.100*** | -0.213*** | 135.9 | 0 | 0.385 | 0.280 |
| Mathematics | 9254 | -0.096*** | -0.196*** | 50.1 | 0 | 0.475 | 0.164 |
| Physics | 37022 | -0.094*** | -0.196*** | 311.4 | 0 | 0.546 | 0.124 |
| Psychology | 3644 | -0.032 | -0.126*** | 130.6 | 0 | 0.479 | 0.202 |
| Overall | 237998 | -0.040*** | -0.108*** | 8216.8 | 0 | 0.521 | 0.213 |

*Baseline: "Cited publication totally_domestic". Controlling for total citations, number of authors and document type*

*Table 12: OLS regression for 2010-2012 Italian publications. Y = Number of citations that Italian publications receive from totally foreign publications (Log transformed), by disciplinary area, excluding self-citations.*

| Area | Number of obs | Cited_pub_prevalently _domestic | Cited_pub_prevalently _international | F | Prob > F | Root MSE | R-squared |
|---|---|---|---|---|---|---|---|
| Art and humanities | 3319 | 0.094*** | 0.057* | 393.0 | 0 | 0.323 | 0.771 |
| Biology | 38323 | 0.166*** | 0.316*** | 8371.5 | 0 | 0.830 | 0.557 |
| Biomedical research | 43877 | 0.227*** | 0.359*** | 12710.8 | 0 | 0.837 | 0.609 |
| Chemistry | 17799 | 0.104*** | 0.151*** | 1026.0 | 0 | 0.886 | 0.402 |
| Clinical medicine | 76384 | 0.249*** | 0.461*** | 18351.1 | 0 | 0.869 | 0.567 |
| Earth and space sciences | 14171 | 0.205*** | 0.428*** | 1497.2 | 0 | 0.831 | 0.503 |
| Economics | 5822 | 0.103*** | 0.159*** | 618.5 | 0 | 0.667 | 0.647 |
| Engineering | 47095 | 0.125*** | 0.230*** | 5223.0 | 0 | 0.845 | 0.481 |
| Law political and social sciences | 5642 | 0.135*** | 0.207*** | 785.5 | 0 | 0.622 | 0.651 |
| Mathematics | 9254 | 0.122*** | 0.211*** | 413.5 | 0 | 0.750 | 0.436 |
| Physics | 37022 | 0.139*** | 0.410*** | 4644.0 | 0 | 0.975 | 0.381 |
| Psychology | 3644 | 0.173*** | 0.300*** | 1114.4 | 0 | 0.777 | 0.594 |
| Overall | 237998 | 0.187*** | 0.417*** | 38718.2 | 0 | 0.915 | 0.483 |

*Baseline: "Cited publication totally_domestic". Controlling for total citations, number of authors and document type*

**4. Discussion and conclusions**

This work analyzes the relation between the level of internationalization of Italian scientific production and that of the relevant citing publications, at an aggregate level and by disciplinary area. The analysis is conducted using bibliometric techniques and as always, the limitations and assumptions embedded in such analyses apply. Caution is therefore recommended in interpreting the findings. The scientific production examined is from the period 2010-2012, with the relevant citing publications observed until 31/05/2017.

The results at aggregate level reveal that more than half of Italian scientific production is totally domestic, while the remaining part is divided almost equally between prevalently domestic and prevalently international. Interestingly, more than half of the citations arrive from totally foreign publications, regardless of the level of internationalization of the cited publications. This finding opens to question the consolidated thesis of the geographical proximity of knowledge spillovers, previously demonstrated through the analysis of patent citations (Jaffe, Trajtenberg, & Henderson, 1993): unlike technical knowledge encoded in patents, scientific knowledge appears to flow rather easily across national borders. Further analyses carried out by the authors (Abramo, D'Angelo, & Di Costa, 2020b) reveal that in domestic knowledge flows, geographic proximity remains an influential factor, although with differences among disciplines and decaying along time. At the same time, the effect of distance on continental flows is modest, and negligible on intercontinental flows.

The factor of the more or less domestic level of a publication clearly has an influence on its interest to foreign scholars, as measured by the level of internationalization of the citing publications. In fact, totally domestic publications receive citations from totally Italian publications to a greater extent than do other categories. The inferential analysis reveals that with respect to a totally domestic publication, a prevalently domestic is cited 13.4% less by totally domestic publications, and a prevalently international 34.3% less, with total citations, number of authors, and document type being equal.

At the moment, we can only surmise whether this phenomenon is more attributable to a lower average quality of totally domestic publications, or to the more country-specific nature of the research problems addressed.

The in-depth analysis for each disciplinary area confirms the evidence of the overall analysis, with minor differences. For all disciplinary areas, the share of totally domestic publications is above 50% (with the sole exception of Physics), with a peak in Art & Humanities (75%), a discipline which is clearly quite country specific. In Physics, the total domestic publications are only 43%, reflective of a research area tending to require international collaborations, often involving very large teams, (i.e. high energy physics, and astrophysics).

Cited publications as a share of the total production increase with the level of internationalization, in all disciplinary areas except Art and humanities, Biomedical research, Economics, and Law, political and social sciences. With regard to citing publications, in all cases, the analyses at the level of the disciplines confirm that as internationalization of the cited publications increases then so does the incidence of totally foreign citing publications.

Totally-Italian citing publications tend to cite more totally domestic publications in the specific areas of Art & Humanities, Earth and space science and Mathematics. The first two of these disciplinary areas are clearly cases where in many subfields the "geographical" context is particularly important. The resulting publications might therefore be of interest to the international scientific community for their methodological content more than for the results themselves, which would be difficult to export or replicate in foreign contexts.

The inferential analysis confirms that, all others equal, as compared to totally domestic publications, prevalently domestic ones receive from 12% to 32% citations more from totally foreign publications. Prevalently international publications receive 84% more citations than totally domestic in Earth and space sciences, +82% in Physics, +79% in Clinical medicine.

To control for the effect of self-citations on the results of the analysis, we have replicated it excluding self-citations from the dataset. What emerges is that, although still statistically significant, the association between nominal variables (category of cited and category of citing publications) weakens, especially in such disciplines as Chemistry, Clinical medicine and Biomedical research.

Since knowledge is cumulative, the question arises as to the extent that the findings observed in the Italian context would differ from those of other countries, which will have different domestic stocks and level of knowledge. In fact, as noted by Abramo, D'Angelo, and Carloni (2019), there is an ever greater chance that new knowledge will stem from domestic research rather than foreign, as: i) the larger is the country in terms of number of researchers and research resources; ii) the more productive is the research system; and iii) the more scientifically advanced the country is in its stock and level of accumulated knowledge, compared to other countries. Given this, it would be interesting to extend this type of analysis to other countries.

A further direction could also be to strengthen the analytical model, particularly in better defining the internationalization of the citing publications, subject to the availability of the author-affiliation link for citing publications. First, the mixed category could be further divided between prevalently domestic and prevalently foreign. Second, the totally foreign category could be distinguished in reference to the bylines, showing either single or multiple countries.

Finally, the Italian academic and research systems have recently been subject to reforms, expected to impact the behavior of organizations and their individual scholars. It would therefore be interesting to extend the analysis to periods subsequent to 2010-2012, to assess whether and how the association between the investigated dimensions has evolved. A specific question, for example, could be whether there has been any change in the incidence of totally Italian publications citing totally domestic publications.